\documentclass[aps,pre,reprint,groupedaddress]{revtex4-1}

\usepackage{amsmath,graphicx}
\usepackage{txfonts}
\renewcommand{\deg}{$^\circ$}

\begin{document}

\title{	Nonlinear shallow ocean wave soliton interactions on flat beaches	}

\author{Mark J.\ Ablowitz}
\author{Douglas E.\ Baldwin}
\email[]{nlwaves@douglasbaldwin.com}
\homepage[]{http://www.douglasbaldwin.com}
\affiliation{Department of Applied Mathematics, University of Colorado, Boulder, Colorado, 80309-0526, USA}

\date{\today}

\begin{abstract}
Ocean waves are complex and often turbulent. While most ocean wave interactions are essentially linear, sometimes two or more waves interact in a nonlinear way. For example, two or more waves can interact and yield waves that are much taller than the sum of the original wave heights. Most of these nonlinear interactions look like an X or a Y or two connected Ys; at other times, several lines appear on each side of the interaction region. It was thought that such nonlinear interactions are rare events: they are not. Here we report that such nonlinear interactions occur every day, close to low tide, on two flat beaches that are about 2,000 km apart. These interactions are closely related to the analytic, soliton solutions of a widely studied multi-dimensional nonlinear wave equation. On a much larger scale, tsunami waves can merge in similar ways.
\end{abstract}



\maketitle


The study of water waves has a long and storied history, with many important applications including naval architecture, oil exploration, and tsunami propagation. 
The mathematics of these waves is difficult because the underlying equations are strongly nonlinear and have a free boundary where water meets air;  there is no comprehensive theory. 
Here we report that X, Y, and more complex nonlinear interactions frequently occur on two widely separated flat beaches and are not rare events, as was previously thought. 
In fact, these X- and Y-type interactions can be seen daily, shortly before and after low tide. 
These phenomena are closely related to the analytical solution of a multi-dimensional nonlinear wave equation that has been studied extensively since 1970 \cite{kp,AbCl91} and is a generalization of an equation studied by Korteweg and de Vries in 1895 \cite{kdv}, which gave rise to the concept of solitons \cite{zabusky:65}. 
From the universality of the underlying equation \cite{Ablowitzbook2011} and the fundamental nature of these waves, it is expected that similar X- and Y-type structures will be seen in many different physical problems, including fluid dynamics, nonlinear optics, and plasma physics. 

\section*{Background and introduction}

Water waves have been studied by mathematicians, physicists, and engineers for hundreds of years. 
While there are many types of water waves, here we will discuss solitary waves in shallow water; 
they are often called solitons and they have unique properties. 
Solitary waves in fluids \cite{Grimshaw2007} and oceans \cite{Osborne2010} are a major and active research area. 

J.\ S.\ Russell, a naval architect, made the first recorded observation of a solitary wave in the Union Canal, Edinburgh in 1834: 
a stopping barge set off a solitary wave that went along the canal for one or two miles without changing its speed or its shape \cite{Russell}. 
He did experiments and found, among other things, that the wave's speed depends on its height; so he concluded that it must be a nonlinear effect. 
J.\ Boussinesq \cite{Boussinesq} in the 1870s and D. Korteweg and his student G. de Vries \cite{kdv} in 1895 derived approximate nonlinear equations for shallow water waves. 
They found both solitary and periodic nonlinear wave solutions to these equations; 
they also found that the speed is proportional to its amplitude --- bigger waves move faster.
So Russell's observations were quantitatively confirmed. 

Between 1895 and 1960, solitary waves were mostly studied by water wave scientists, mathematicians, and coastal engineers. 
In the 1960s, applied mathematicians developed robust approximation techniques and found that the Korteweg--de Vries (KdV) equation appears universally when there is weak quadratic nonlinearity and weak dispersion \cite{Ablowitzbook2011}. 
In 1965, Zabusky and Kruskal \cite{zabusky:65} found that the solitary waves of the KdV equation have remarkable elastic interaction properties and termed them solitons. 
Gardner, Greene, Kruskal, and Miura \cite{GaGeKrMi67} then developed a method for solving the KdV equation with rapidly decaying initial data;
this method has been extended to many other nonlinear equations and is called the Inverse Scattering Transform (IST) \cite{Ablowitz1981IST,NMPZ84} --- such equations are called integrable. 

In 1970, Kadomtsev and Petviashvili \cite{kp} (KP) extended the KdV equation to include transverse effects; 
this multi-dimensional equation, like the KdV equation, is integrable \cite{AbCl91}. 
Our observations in this article are related to soliton solutions of the KP equation that do not decay at large distances; 
these interacting, multi-dimensional line soliton solutions can be found analytically \cite{Ablowitz1981IST}. 
Before our observations, there was only one well-known photograph of an interacting line soliton in the ocean and it was thought that such interactions are rare events; 
it was taken in the 1970s in Oregon (Fig.~4.7b in ref \cite{Ablowitz1981IST}) and is similar to Fig.~\ref{fig:mediumX}. 
Since the KP equation has other X, Y, and more complex line soliton solutions, we sought and found ocean waves with similar behavior (Figs.~\ref{fig:shortX}--\ref{fig:3in2out}). 
Surprisingly, these X, Y, and more complex types of line solitons appear frequently in shallow water on two relatively flat beaches, some 2,000~km apart! 
These freely propagating, interacting line solitons are remarkably robust. 
While these interactions are not stationary and so only last a few seconds, a casual observer will be able to see them with the insights provided in this article. 
Interestingly,  in laboratory experiments involving internal waves emanating from the interaction of cylindrical wave fronts, \citet[Fig.~11]{Maxworthy1980} reported an X-type internal wave interaction; \citet{Weidman1992} later showed that the length of the stem in \cite[Fig.~11]{Maxworthy1980} follows a Hopf bifurcation when plotted against the intersection angle.

\section*{Observations}

\begin{figure}[t]
	\centering
	\includegraphics{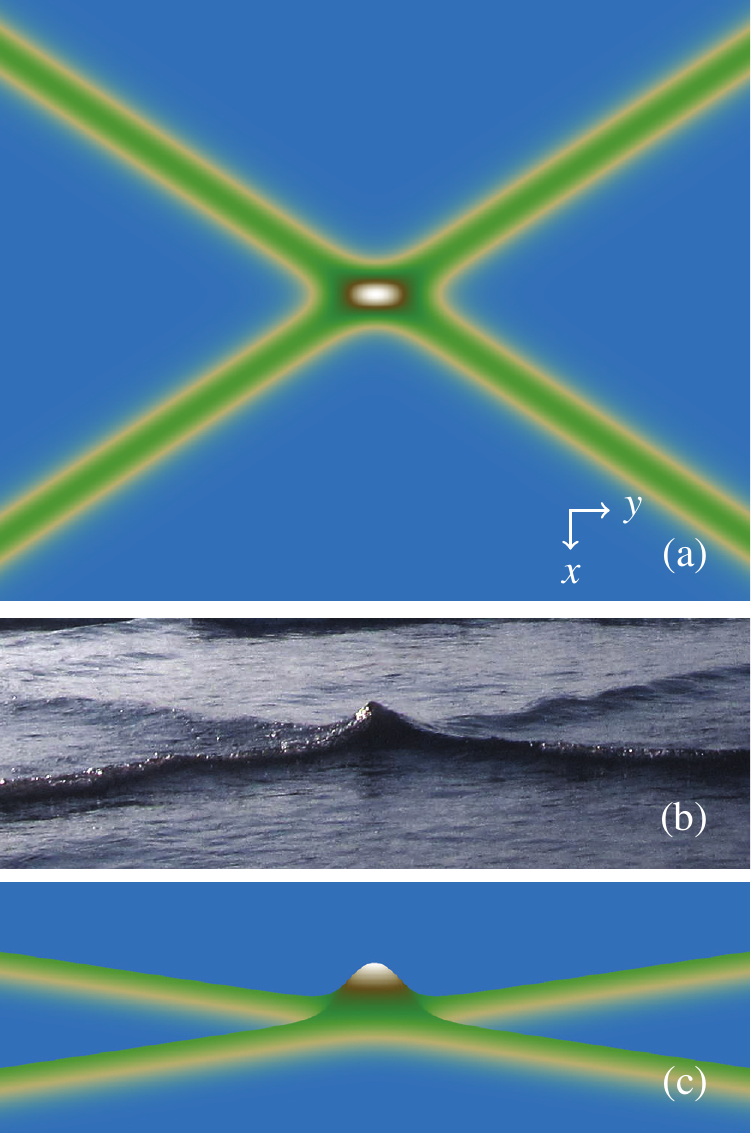}
	\caption{\label{fig:shortX}
		A plot and a photograph of an X-type interaction. 
		(a)~A plot of an analytical line-soliton interaction solution of the KP equation at $t=0$ using \eqref{eq:kpdirect} and \eqref{eq:theF}. 
		In this and the following plots, we picked the $k_i$ and $P_j$ to be qualitatively similar to the photograph in part~(b).
		Here, $k_1=k_2=1/2$, $P_1=-P_2=2/3$ so $e^{A_{12}} \approx 2.3$.
		(b)~Taken in Mexico on 31 December 2011; notice the large amplitude of the short stem.
		(c)~A 3d-plot of the solution in (a), which qualitatively agrees with (b); we only include one 3d-plot because the density plots show the interaction behavior clearly.
	}
\end{figure}

\begin{figure}[t]
	\centering
	\includegraphics{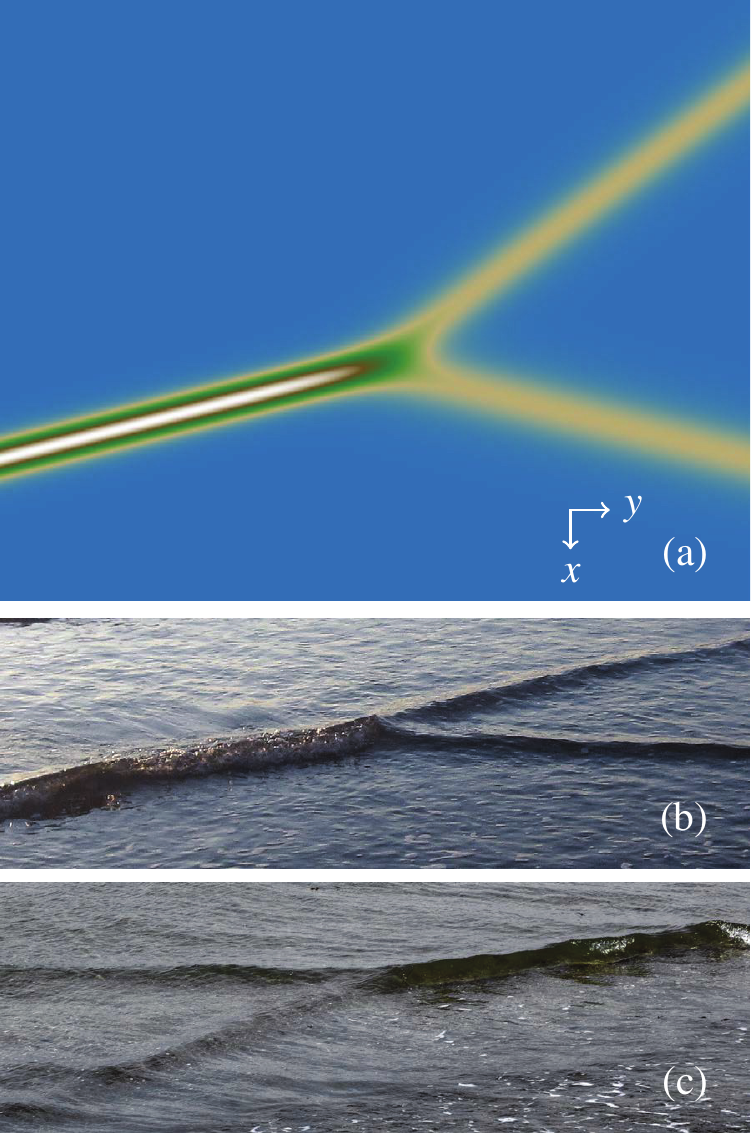}
	\caption{\label{fig:Y}
		A plot and photographs of a Y-type interaction.
		(a)~$k_1=1/2$, $k_2=1$, $P_1=3/4$, $P_2=1/4$ so $e^{A_{12}} = 0$.
		(b)~Taken in Mexico on 6 January 2010.
		(c)~Taken in California on 3 May 2012.
	}
\end{figure}

\begin{figure}[t]
	\centering
	\includegraphics{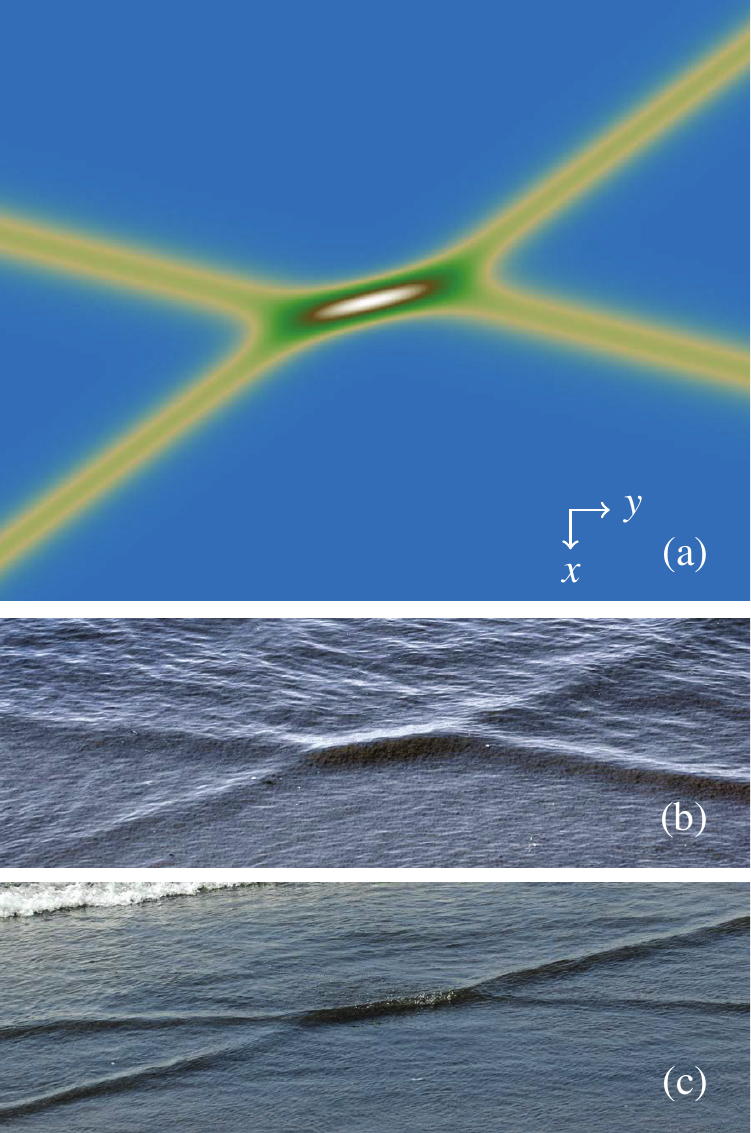}
	\caption{\label{fig:mediumX}
		A plot and photographs of an X-type interaction with a longer stem.
		(a)~$k_1=k_2=1/2$, $P_1=-1/4-10^{-2}$, $P_2=3/4$ so $e^{A_{12}} \approx 51$.
		(b)~Taken in California on 2 May 2012 in shallower water than Fig.~\ref{fig:shortX}b.
		(c)~Taken in California on 4 May 2012.
	}
\end{figure}

\begin{figure}[t]
	\centering
	\includegraphics{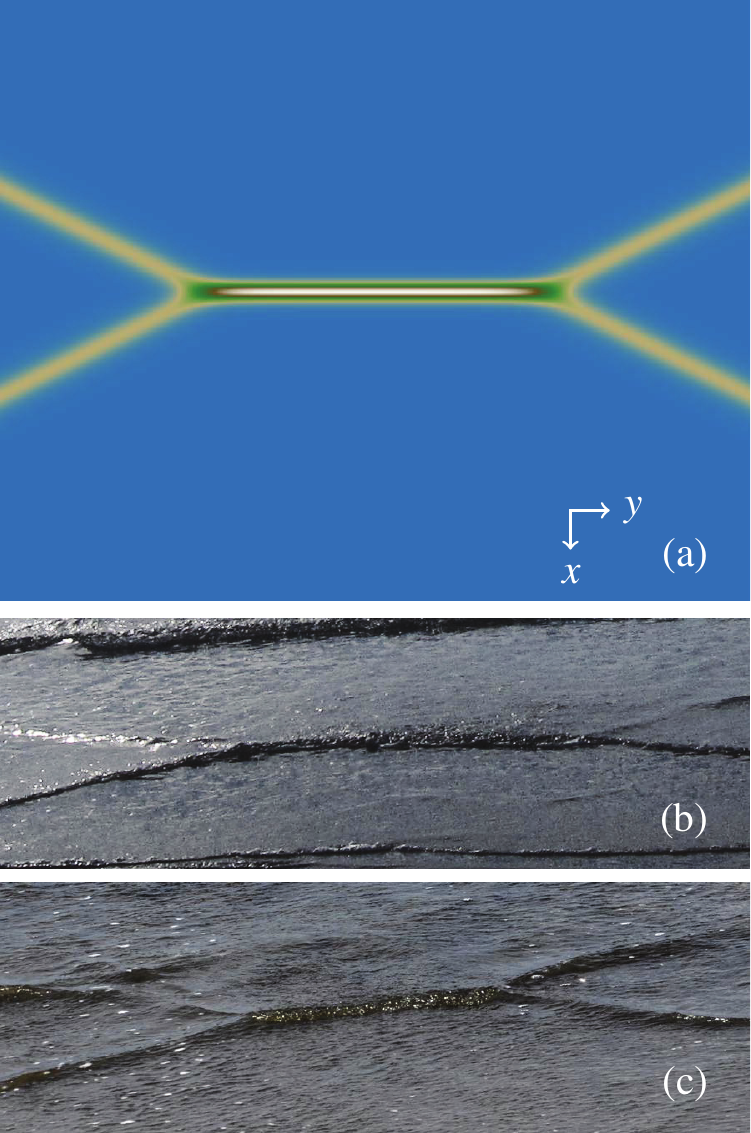}
	\caption{\label{fig:longX}
		A plot and photographs of an X-type interaction with a very long stem.
		(a)~$k_1=k_2=1/2$, $P_2=-P_1+10^{-10}=1/2$ so $e^{A_{12}} \approx 5\times10^{9}$.
		(b)~Taken in Mexico on 28 December 2011 in shallower water than Fig.~\ref{fig:mediumX}b.
		(c)~Taken in California on 3 May 2012.
	}
\end{figure}

\begin{figure}[t]
	\centering
	\includegraphics{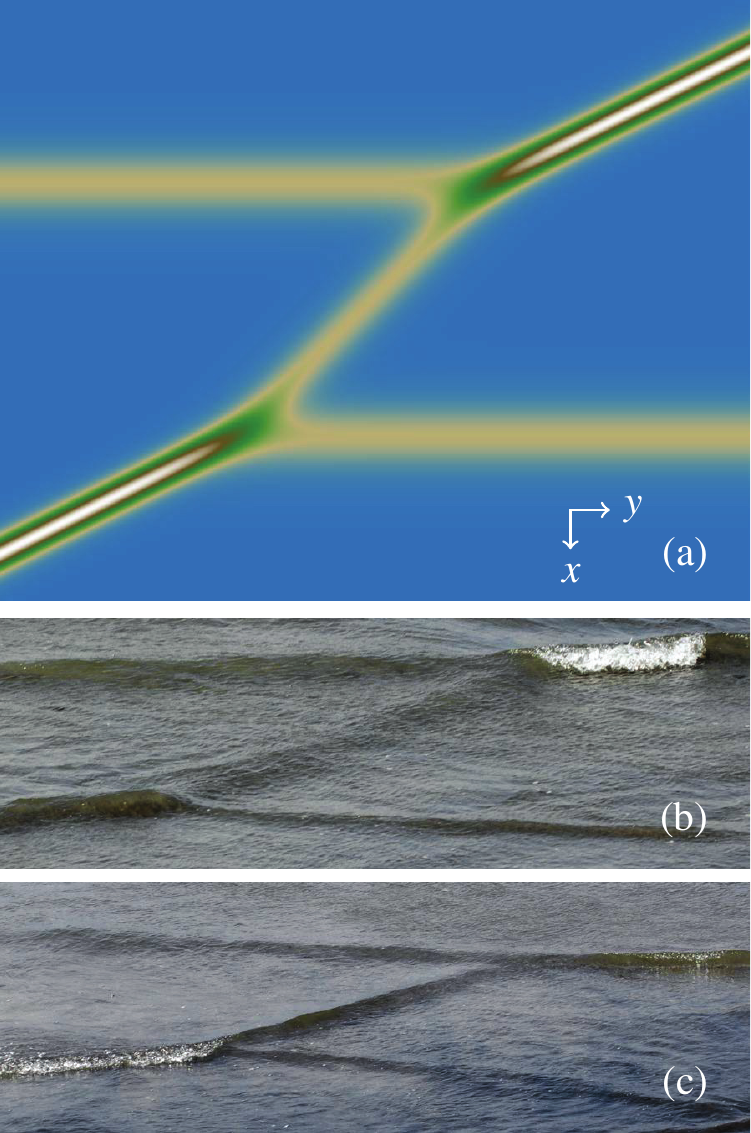}
	\caption{\label{fig:neglongX}
		A plot and photographs of an X-type interaction, where the stem has a lower rather than a higher amplitude.
		(a)~$k_1 = 1$, $k_2 = 1/2$, $P_1 = 1/2-10^{-7}$, $P_2 = 0$ so $e^{A_{12}} \approx 5\times10^{-8}$.
		(b) and (c) Taken in California on 3 May 2012.
	}
\end{figure}

\begin{figure}[t]
	\centering
	\includegraphics{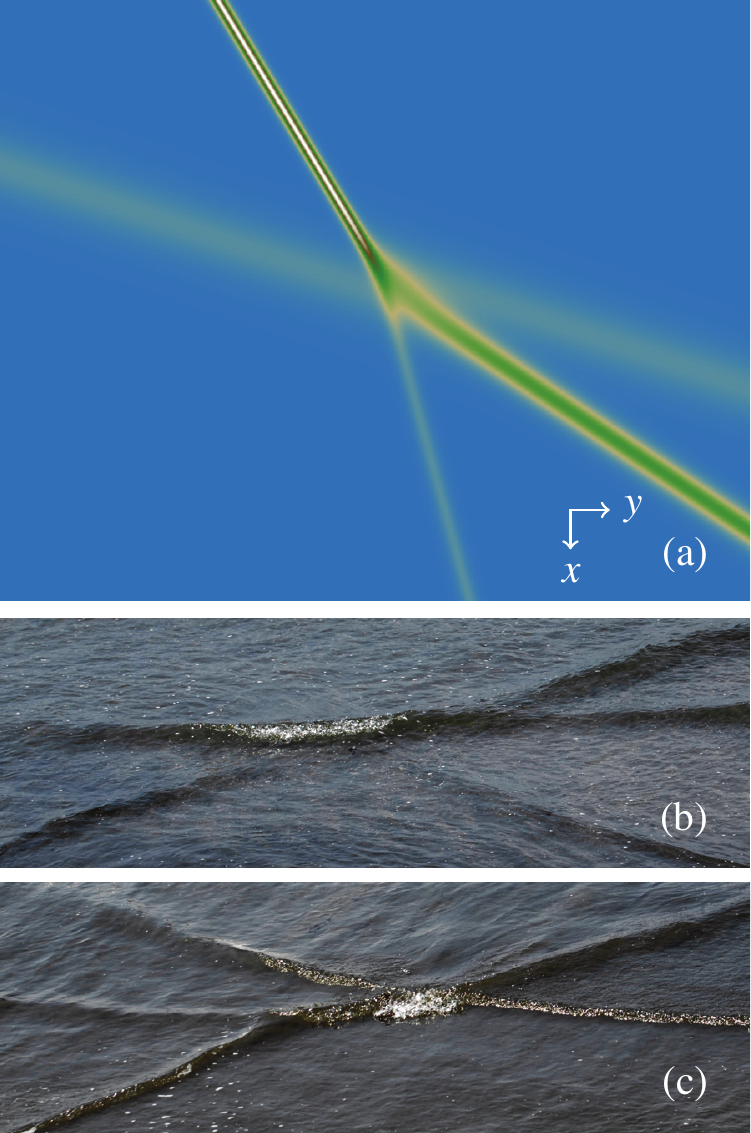}
\caption{\label{fig:3in2out}
	A plot and photographs of a 3-in-2-out interaction, where there are three line solitons on one side of the interaction region and two line solitons on the other side. 
	(a)~$k_1 = 1$, $k_2 = 2$, $k_3 = 3$, $P_1 = -1/3$, $P_2 = -2/3$, $P_3 = -5/3$.
	(b) and (c)~Taken in California on 4 May 2012.
	}
\end{figure}

Single line, solitary water waves are familiar to every beach goer: 
they are localized in the direction of propagation and have a distinctive, hump-like wave profile. 
These waves break when they are sufficiently large compared to the depth and they often curve from transverse beach and bottom effects. 
We will focus on interacting line solitary waves that form X, Y, and more complex interactions. 

It was thought that X-type ocean wave interactions happen infrequently. 
This is not the case: 
X- and Y-type ocean wave interactions occur daily, shortly before and after low tide on relatively flat beaches. 
M.J.A. observed these interactions near 20\deg 41'22"N, 105\deg 17'44"W in Nuevo Vallarta, Mexico from 2009 to 2012 between December and April. 
D.B. observed these interactions near 33\deg 57'52"N, 118\deg 27'35"W on Venice Beach, California in May 2012 --- about 2,000~km away.
Figs.~\ref{fig:shortX}--\ref{fig:3in2out} shows a few of the thousands of photographs that we took. 
The water depth where we saw these interactions was shallow, usually between 5 and 20~cm; 
the beaches are long and relatively flat; 
the interactions usually happen within 2~hours before and after low tide; 
the cross-waves produced near a jetty appear to help induce these interactions.
We found that these X- and Y-type interactions usually come in groups, which last a few minutes. 
We saw many X- and Y-type interactions each day that we made observations; 
the relative frequencies of the interactions were different at the two beaches --- M.J.A. saw X-type interactions like Fig.~\ref{fig:shortX} more often than D.B. 
We also saw more complex interactions, such as three line solitons on one side of the interaction region and two line solitons to the other side, which we will call a 3-in-2-out interaction; 
these more complex interactions are much less frequent than X- and Y-type interactions. 
Our observations indicate that X- and Y-type interactions are remarkably robust: 
they typically persist through bottom-depth changes, perturbations from wind and spray, and sometimes even breaking! 

We observed three types of X interactions: 
an interaction with a short stem (Fig.~\ref{fig:shortX}), 
an interaction with a long stem where the stem height is higher than the incoming line solitons (Figs. \ref{fig:mediumX} and \ref{fig:longX}), and 
an interaction with a long stem where the stem height is lower than the tallest incoming line soliton (Fig.~\ref{fig:neglongX}). 
The amplitude of the short-stem X-type interaction can be quite large in deeper water. 
Interestingly, the length of the stem often increases as the depth decreases. 
Fig.~\ref{fig:Y} shows a typical Y-type interaction. 
A more complex interaction, with three `incoming' and two `outgoing' segments, is shown in Fig.~\ref{fig:3in2out}. 

When one knows what to look for and when and where to look for them,  X- and Y-type interactions are fairly easy to observe. 
In addition to happening less frequently, more complex interactions are harder to see because they are highly non-stationary and have shorter interaction times than X- and Y-type interactions. 
Another difficulty is that most water waves break before X- or Y-type interactions form; 
so sustained observation may be needed. 
Along with the photographs here, we have also taken many videos that show the development and general dynamics of these waves; 
the readers can watch some of these videos and see many more photographs at our websites \cite{AblowitzBaldwinWeb}. 

\section*{Mathematical description}

The KP equation \cite{kp}, 
\begin{equation} \label{eq:kp_dim}
	\frac{\partial}{\partial x}
	 \left( \frac{1}{\sqrt{gh}} \eta_t + \eta_x 
		+ \frac{3}{2h} \eta\eta_x + \frac{h^2\gamma}{2}\eta_{xxx} \right)
		+ \frac{1}{2} \eta_{yy} = 0,
\end{equation}
is the two-space and one-time dimensional equation that governs unidirectional, maximally-balanced, weakly-nonlinear shallow water waves with weak transverse variation. 
Here, sub-scripts denote partial derivatives, 
$\eta =\eta(x,y,t)$ is the wave height above the constant mean height $h$, 
$g$ is gravity,
$\gamma = 1-\tau/3$, 
$\tau = T/(\rho g h^2)$ is a dimensionless surface tension coefficient, 
and $\rho$ is density. 
When there is no $y$-dependence, the equation reduces to the KdV equation \cite{kdv}. 
The KP equation was first derived in the context of plasma physics \cite{kp} and was later derived in water waves \cite{Ablowitz1979}. 
The sign of $\gamma$ is important: 
there is `large' surface tension when $\gamma<0$ and this equation is called KPI; 
there is `small' surface tension when $\gamma>0$ and this equation is called KPII. 
We can rescale \eqref{eq:kp_dim} into the non-dimensional form \cite{Ablowitzbook2011}
\begin{equation} \label{eq:kp_nondim}
 ( u_t + 6 u u_x + u_{xxx} )_x + 3\sigma u_{yy} = 0,
\end{equation}
where $u$ relates to the wave height $\eta$ 
and $\sigma=\pm1$ corresponds to the sign of $\gamma$.

For large surface tension, KPI has a lump-type solution that decays in both $x$ and $y$ but has not yet been observed. 
Only recently has a large-surface-tension one-dimensional soliton been observed \cite{Falcon2002}; it satisfies the KdV equation and is a depression from the mean height.

We will only discuss KPII here because surface tension is small for ocean waves. 
The KPII equation has solutions with a single-phase, which we will call line-solitons.
We are interested in the interactions of line solitons. 
These solutions can be found by so-called direct methods \cite{Ablowitz1981IST}:  
special $N$-soliton solutions of the KP equation can be written in the form \cite{Satsuma1976}
\begin{equation}  \label{eq:kpdirect}
	u=u_N= 2 \frac{\partial^2 F_N}{\partial x^2},
\end{equation}
where $F_N$ is a polynomial in terms of suitable exponentials. 
This solution is convenient for finding the simplest such solution: 
the first three are 
\begin{subequations} \label{eq:theF}
\begin{equation}
	\begin{gathered}
	F_1 = 1 + e^{\eta_1}, \qquad
	F_2 = 1 + e^{\eta_1} + e^{\eta_2} + e^{\eta_1+\eta_2+A_{12}}, \\
	F_3 = 1 + \sum_{1 \le i \le 3} e^{\eta_i}
		+ \sum_{1 \le i < j \le 3} e^{\eta_i + \eta_j + A_{ij}}
		+ e^{\eta_1 + \eta_2 + \eta_3 + A_{12} + A_{13} + A_{23}},
	\end{gathered}
\end{equation}
where $\eta_j= k_j[x+P_jy-(k_j^2+3 \sigma P_j^2)t]+\eta_j^{(0)}$, 
$k_j$, $P_j$, $\eta_j^{(0)}$ are constants, 
and 
\begin{equation}
	e^{A_{ij}}
	= \frac{(k_i-k_j)^2-\sigma(P_i-P_j)^2}%
		{(k_i+k_j)^2-\sigma(P_i-P_j)^2}, \qquad i < j. 
\end{equation}
\end{subequations}

For KPII (where $\sigma=1$), 
$u_1$, $F_1$ corresponds to the simplest one line soliton, which is essentially one-dimensional. 
The more interesting case of $u_2$, $F_2$ corresponds to the interaction of two line soliton waves.  
These interactions have distinct patterns:  
when $e^{A_{12}} = O(1)$, we get an X-type interaction with a short stem (Fig.~\ref{fig:shortX}); 
when $e^{A_{12}} \gg 1$, we get an X-type interaction with a long stem where the stem height is higher than the incoming line solitons (Figs. \ref{fig:mediumX} and \ref{fig:longX}); 
when $e^{A_{12}} \ll 1$, we get an X-type interaction with a long stem where the stem height is less than the height of the tallest incoming line soliton (Fig.~\ref{fig:neglongX}); and
when $e^{A_{12}} = 0$, we get a Y-type interaction (Fig.~\ref{fig:Y}). 
As mentioned earlier, the length of the stem appears to be correlated to the depth of the water. 
Short stems where $e^{A_{12}} = O(1)$ are usually found in much deeper water than long stem X- or Y-type wave interactions where $e^{A_{12}} \gg 1$ or $e^{A_{12}} \ll 1$.

Recently, novel and exotic web-like structures for the KP equation ($N$-in-$M$-out) have been found using Wronskian methods \cite{BK03,CK08} that go beyond the simplest `building block' solutions of X- and Y-type line soliton solutions. 
Note also that an $N$-in-$M$-out solution (where $M<N$) can be found by starting with $F_N$ and taking $k_i$ and $P_j$ such that $e^{A_{M,N}} = \dotsb = e^{A_{N-1,N}} = 0$; 
Fig.~\ref{fig:3in2out} shows such a 3-in-2-out interaction. 
It was recently shown that these line interactions persist under the next order perturbations in the equations for water waves \cite{abcurtis2011}; 
while the stem can be four times the height of the incoming line solitons in the KP equation, it is less than four times the height when higher-order terms are included. 

\section*{X- and Y-type structures and tsunami propagation}

Miles \cite{milesone,milestwo} first discovered that Y-type solutions could be associated with the KP equation; 
he also related it to ``Mach-stem reflection'', the phenomenon that occurs in gas dynamics. 
Interestingly, Wiegel \cite{weigel} reported that the 1946 Aleutian earthquake induced tsunami caused a Mach-stem reflection along the cliffs of the western edge of Hilo Bay in Hawaii. 
Yeh et al. \cite{yeh} revisited Mach-stem reflection in water waves with an inclined bottom, both analytically in the context of the KP equation and in a laboratory water wave tank. 

Recent observations of the 2011 Japanese Tohoku-Oki earthquake induced tsunami indicate that there was a `merging' phenomenon from a cylindrical-wave-type interaction \cite{Song2012} that significantly amplified the tsunami and its destructive power. 
This effect is remarkably similar to the initial formation of an X- or Y-type wave:
while it is initially a linear super-position effect, 
the interaction can be significantly modified or enhanced by nonlinearity after propagating to shore. 
Moreover, for large distances (in the open ocean direction) an earthquake induced tsunami will propagate approximately like the KP equation. 
So strong nonlinear effects from X- or Y-type interactions can have serious effects for land much further away;  
the destruction in Sri Lanka from the 2004 Sumatra--Andaman earthquake induced tsunami is an example of such a long distance effect.

\section*{Conclusion} 

We reported that X- and Y-type shallow water wave interactions on a flat beach are frequent, not rare, events. 
Casual observers can see these fundamental wave structures once they know what to look for. 
Extensive ocean observations reported here enhance and complement laboratory and analytical findings. 
We expect that similar interactions will be observed in many other fields --- including fluid dynamics, nonlinear optics, and plasma physics --- because the leading-order equation here is also the leading-order equation for many other physical phenomena.

\section*{Acknowledgements}

We wish to acknowledge the support of the National Science Foundation under grant DMS-0905779.

\bibliographystyle{apsrev}
\bibliography{references}

\end{document}